\PassOptionsToPackage{sort&compress,numbers,super}{natbib}
\documentclass[%
    aip, jcp,%
    letterpaper,
    amsmath,amssymb,%
    reprint,%
]{revtex4-1}

\usepackage{titlesec}
\titleformat{\section}
    {\sffamily\large\bfseries}{}{1em}{\MakeUppercase}
\titleformat{\subsection}[runin]
    {\sffamily\bfseries}{}{0pt}{}[.]
\titlespacing*{\subsection}
{0pt}{0.5\parskip}{0.5em}

\usepackage[journal=ancac3, manuscript=article, layout=twocolumn]{achemso}
\usepackage{natbib}
\usepackage{natmove}
\setcitestyle{sort&compress,numbers,super}

\usepackage{graphicx}

\usepackage[utf8]{inputenc}
\usepackage[T1]{fontenc}
\usepackage{mathptmx}

\usepackage{siunitx}

\usepackage{booktabs}
\usepackage{tabularx}
\usepackage{ragged2e} 

\newcommand{\pder}[2][]{\ensuremath{\dfrac{\partial#1}{\partial#2}}}
\newcommand{\dtime}{\ensuremath{\partial_t}}

\newcommand{\bvec}[1]{\mathbf{#1}}
\newcommand{\ncdot}{\!\cdot\!}
\newcommand{\dd}{\mathrm{d}}
\newcommand{\kT}{\ensuremath{k_\text{B}T}}

\newcommand{\dpart}{\ensuremath{d}}
\newcommand{\mob}{\mu}
\newcommand{\tsub}[1]{\ensuremath{_\text{#1}}}
\newcommand{\zR}{z_\text{R}}
\newcommand{\nliq}{\ensuremath{n\tsub{liq}}}

\newcommand{\Fgrad}{\ensuremath{\bvec F_{\!\nabla}}}
\newcommand{\Fscat}{\ensuremath{\bvec F_\text{\!pr}}}
\newcommand{\Fopt}{\ensuremath{\bvec F_\text{\!opt}}}
\newcommand{\Pthresh}{\ensuremath{P_\text{thresh}}}
\newcommand{\crossec}[1]{\ensuremath{\sig\tsub{#1}}}
\newcommand{\NA}{\operatorname{\mathrm{N\kern-0.1em A}}}
\newcommand{\textNA}{{N\kern-0.1em A}}
\newcommand{\psource}{\ensuremath{S_{\!c}}}
\newcommand{\psourced}{\ensuremath{S_{\!d}}}
\newcommand{\psourcefrac}{\ensuremath{s_{d}}}
\newcommand{\pflux}{\ensuremath{j}}
\newcommand{\qdepo}{\ensuremath{q\tsub{depo}}}
\newcommand{\fluxIntBound}{\ensuremath{\Gamma^{-}_{\!\pflux}}}
\newcommand{\flpress}{\ensuremath{P}}

\newcommand{\Pec}{\operatorname{\mathit{P\kern-0.14em e}}}
\newcommand{\Rey}{\operatorname{\mathit{R\kern-0.08em e}}}

\newcommand{\lam}{\lambda}
\newcommand{\sig}{\sigma}

\newcommand{\unit}[1]{\,\si{#1}}
\newcommand{\ttenTo}[1]{\ensuremath{\ncdot 10^{#1}}}

\newcommand{\ITWM}{Fraunhofer Institute for Industrial Mathematics ITWM, 67663 Kaiserslautern, Germany}
\newcommand{\TUK}{Physics Department and State Research Center OPTIMAS, Technische Universit\"at Kaiserslautern, 67663 Kaiserslautern, Germany}

\begin{document}

\title{Convection, Heat Generation and Particle Deposition in Direct Laser Writing of Metallic Microstructures}

\author{Thomas Palmer}
    \email{thomas.palmer@itwm.fraunhofer.de}
    \affiliation{\ITWM}

\author{Erik H.\ Waller}
    \affiliation{\TUK}

\author{Heiko Andr\"a}
    \affiliation{\ITWM}

\author{Konrad Steiner}
    \affiliation{\ITWM}

\author{Georg von Freymann}
    \affiliation{\TUK}
    \affiliation{\ITWM}

\date{\today}

\begin{abstract}
    
Three-dimensional metallic microstructures find applications as stents in medicine, as ultrabroadband antennas in communications, in micromechanical parts or as structures of more fundamental interest in photonics like metamaterials.
Direct metal printing of such structures using three-dimensional laser lithography is a promising approach, which is not extensively applied yet, as fabrication speed, surface quality, and stability of the resulting structures are limited so far.
In order to identify the limiting factors, we investigate the influence of light-particle interactions and varying scan speed on heat generation and particle deposition in direct laser writing of silver.
We introduce a theoretical model which captures diffusion of particles and heat as well as the fluid dynamics of the photo-resist.
Chemical reactions are excluded from the model but particle production is calibrated using experimental data.
We find that optical forces generally surmount those due to convection of the photo-resist.
Simulations predict overheating of the photo-resist at laser powers similar to those found in experiments.
The thermal sensitivity of the system is essentially determined by the largest particles present in the laser focus.
Our results suggest that to improve particle deposition and to achieve higher writing speeds in metal direct laser writing, strong optical trapping of the emerging particles is desirable.
Furthermore, precise control of the particle size reduces the risk of spontaneous overheating.

\end{abstract}

\maketitle


\begin{figure*}
    \includegraphics[width=0.6\textwidth]{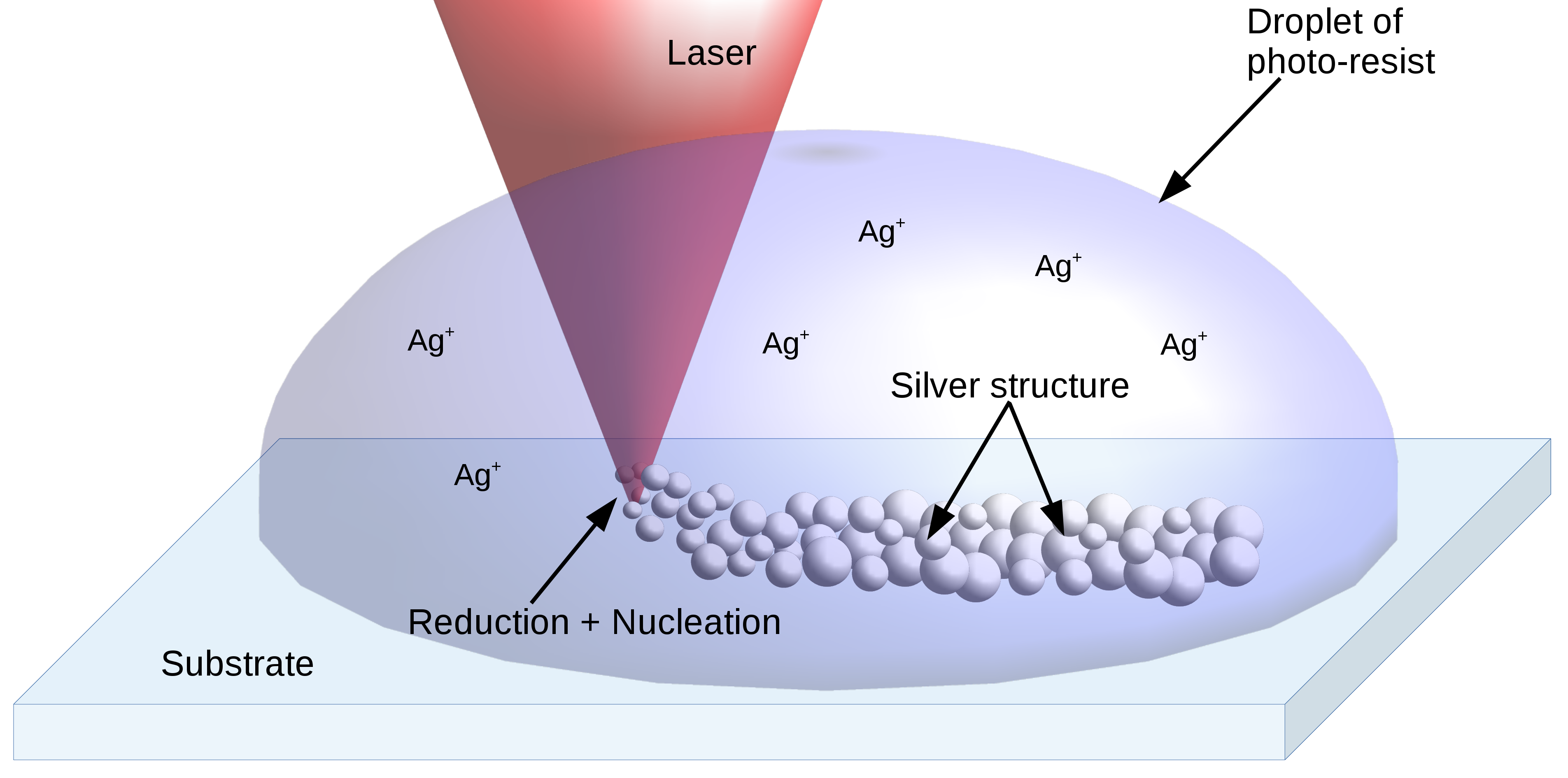}
    \qquad
    \includegraphics[width=0.25\textwidth]{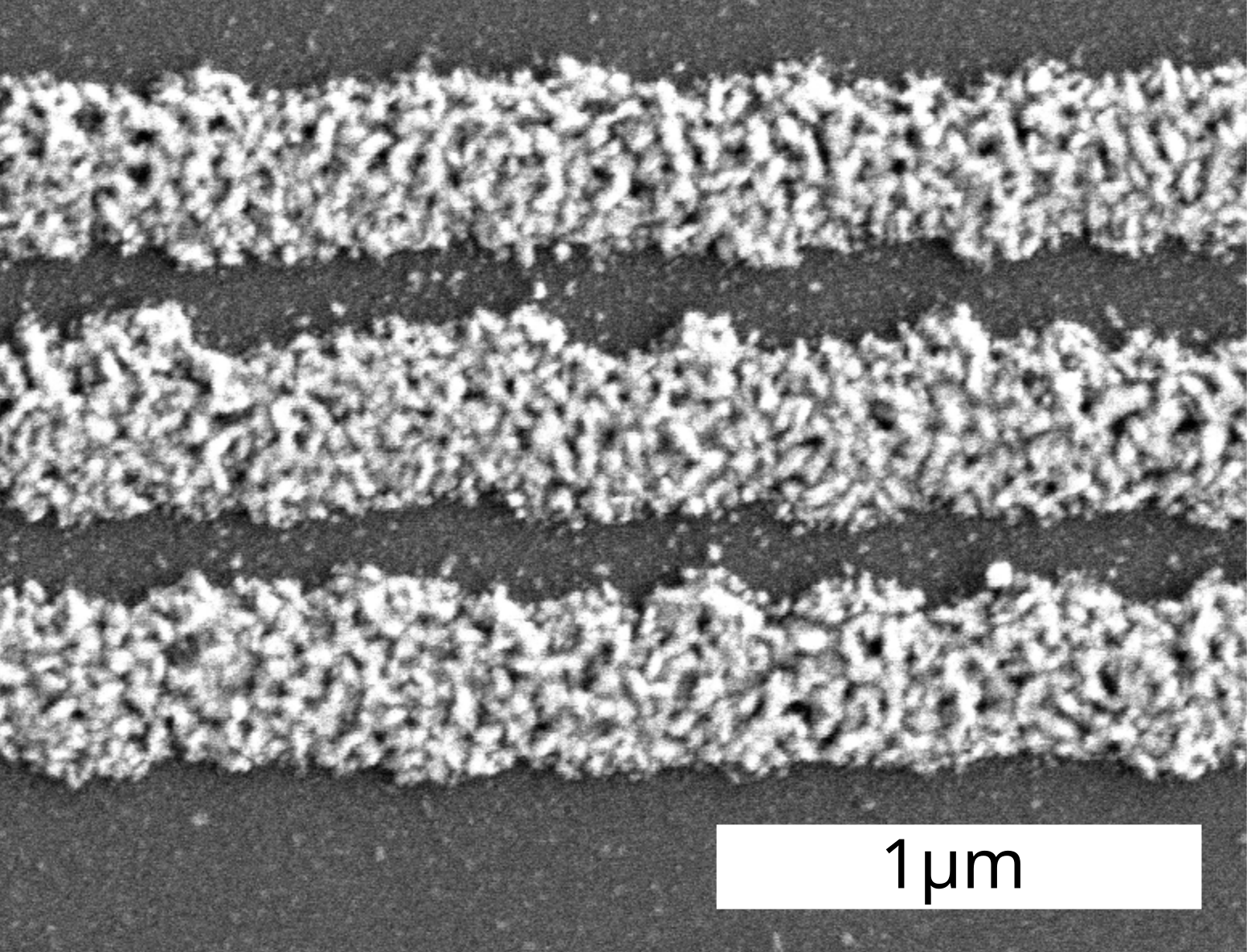}
    \caption{Left: Illustration of MDLW process:
        a pulsed laser beam is tightly focussed into a photo-resist,
        this locally initiates a photo-reaction which ultimately results in a small solid portion of the final structure.
        Scanning the laser focus on a 2D or 3D trajectory, the desired structure is written.
        Right: SEM image of lines written with MDLW.
        The granularity indicates that the structure consists of silver nano particles that agglomerate.
    }
    \label{fig:fancyPlot}
\end{figure*}

Direct laser writing of metallic micro-structures (MDLW) emerges as a fabrication method\cite{Waller2021}
complementing the established process of polymer-based direct laser writing (DLW).
DLW is proven to yield highly sophisticated 3D polymeric microstructures
and, hence, is ideally suited for, \latin{e.g.}, micro-optical applications\cite{Thiele2017, Thiele2019, Toulouse2021, Schumann2014}.
Structures produced using DLW can serve as template for metallic structures\cite{Deubel2004, Rill2008, Gansel2009, Gansel2012},
yet, such fabrication methods limit geometric freedom or require electrically conductive substrates.
MDLW enables the \emph{direct} fabrication of metallic structures on almost arbitrary substrates.
Structures produced via MDLW exhibit an electrical conductivity similar to that of the bulk medium\cite{Tanaka2006},
which gives them an advantage over fabrication methods in which metal nano-particles are initially
embedded into polymers\cite{Blasc2016} or ligands\cite{Greer2007,Greer2018}.

MDLW has been used for fabrication of microelectronics \cite{Waller2019, Waller2018},
micro-sensors \cite{Lee2017} in micro-fluidic channels and writing on flexible substrates \cite{He2017c}.
To improve fabrication of more sophisticated structures, e.g. photonic meta-materials or micro-optical devices,
higher fabrication speeds and smoother surface finish 
are desirable.
In addition, the growing nano-particles partially absorb the incident laser light and heat the surrounding photo-resist,
which can lead to evaporation and disruption of the writing process.
Key to these challenges is the strong laser-particle interaction\cite{Waller2019}
whose role in the MDLW process we aim to better understand in order to increase fabrication speed as well as structural quality.

The underlying effects that determine the above mentioned phenomena have individually been investigated in depth,
both experimentally and by computer simulations,
\latin{e.g.}, plasmonic heating\cite{Baffo2009, Baffo2010, Baffo2011, Baffo2013} and induced fluid convection\cite{Donner2011},
particle trapping\cite{Ashkin_1992, Ashkin_2000, Svobo1994, Nieminen2007, Nieminen2014} and maximum translation velocities of nano-particles\cite{Melzer2018}.
However, the above studies only treat quasi-equilibrium situations where the number of particles is fixed and the laser in rest\cite{Donner2011, Baffou2010, Baffo2013},
or, only a single particle in a moving laser focus is considered\cite{Melzer2018}.
For the competing micro fabrication process of Direct Ink Writing with on-the-fly annealing\cite{Skylar2016},
the temperature evolution has been addressed with simulations.
In a continuous particle generation-deposition process as is the case in MDLW,
diffusive spreading and deposition of particles as well as heating and convection of the photo-resist all depend on each other and on the laser scan speed.
In this paper, we propose a simulation model which depicts these effects.
Thereby, we investigate the interplay of optical configuration and scan speed on the writing process
and characterize the relative importance of the various phenomena involved in the process.

We set off from an idealized source of particles, located at the laser focus.
Since most commonly silver is used in MDLW, we only consider silver structures and nano-particles in this paper,
though our model is not restricted to silver and can be applied to other materials straightforwardly.

This paper is organized as follows:
After revisiting optical trapping of nano-scale particles, we introduce our simulation model.
Then we present results for the convection velocity of the photo-resist
and the temperature increase as a function of laser power.
We analyze the effect of the particle size on the heat generation within the fluid.
Then, we derive the laser powers corresponding to the overheating threshold and compare them to experimental findings.
In the last section, we demonstrate, how optical forces can improve the deposition of particles in MDLW,
and the analysis of experimentally measured particle production rates is shown.
We conclude with a few remarks concerning possible refinements of the simulations and improvements of the MDLW process.
Trailing is a method section where experimental methods and simulations are described in further detail;
numeric values of the model parameters are given and boundary conditions are specified and motivated.

\section{Results and Discussion}

\subsection{Laser particle trapping}
\label{sec:theory}
Reduction and nucleation of silver starts in the focus of the laser beam,
where the generated nano-particles are subject to the electrical field $\bvec E\tsub{ext}$ of the incident light.
The fields induced in the particles by $\bvec E\tsub{ext}$ interact with the latter, resulting in a net force on the particles.
Most important is the induced dipole moment $\bvec p$ which causes a force
\begin{equation}
    \bvec F_\nabla = \nabla (\bvec p \cdot \bvec E\tsub{ext}) = \frac{\nliq \alpha}{2 c_0} \nabla I
    \label{eq:gradForce}
\end{equation}
proportional to the gradient of the irradiance $I$.
Here $c_0$ is the speed of light in vacuum.
The polarizability $\alpha = 3 V_i \frac{m^2 -1}{m^2 + 2}$ depends on the particle's
volume $V_i$ and the ratio of refractive indices $m=n/\nliq$ of particle and surrounding media\cite{BohrenHuffman1983}, respectively.
Eq.\ \eqref{eq:gradForce} describes a conservative force field with potential $\propto -I$.

For the cross-section $\crossec{pr}$ corresponding to the force due to radiation pressure
(a.k.a. phase gradient force, scattering force),
we use the result obtained from Mie-theory\cite{BohrenHuffman1983}.
Then we can determine the force due to radiation pressure
\begin{equation}
    \label{eq:scatForce}
    \bvec F\tsub{pr}
    = (\nabla \Phi) \frac{1}{\omega} \crossec{pr} I
\end{equation}
where $\Phi$ denotes the total phase and $\omega$ is the angular frequency of the electrical field.
Radiation pressure becomes more and more important with growing particle size;
eventually it surmounts the attractive gradient force \eqref{eq:gradForce},
so that larger particles cannot be confined but are ejected from the focus along the beam axis.
For the total optical force we write $\Fopt \equiv \Fgrad + \Fscat$.

\subsection{Model equations}
In case of equilibrium and a conservative force field, any initial
particle distribution will relax into the corresponding Boltzmann-distribution.
In our case, however, the laser moves through the photo-resist and continuously produces silver particles.
This requires to solve the transport equation
\begin{equation}
    \label{eq:particleDiffusion}
    \dtime c + \nabla \cdot \big\{ (\bvec u + \mob \Fopt ) c \big\} = {} - \nabla\cdot(D\nabla c) + \psource
\end{equation}
for the local particle concentration $c$.
The term $(\bvec u + \mob \Fopt) c$ captures advection of particles due to optical forces $\Fopt$ and the local velocity $\bvec u$ of the surrounding photo-resist,
taken in the reference frame of the laser.
Here, particle diameter $d$ and fluid viscosity $\eta$ determine the mobility $\mob = (3\pi \eta \dpart)^{-1}$.
Diffusion is assumed to be Fickian with diffusion constant given by the Einstein-Smoluchowski relation\cite{Gillespie2012}
$D = \mu \kT_0$ where $T_0$ is the ambient temperature and $k\tsub{B}$ the Boltzmann constant.
In a reference frame centered around the moving laser focus,
particle production is represented by the spatially steady source term
\begin{equation}
    \label{eq:particleSource}
    \psource(x,y,z) \propto \exp\left\lbrace - 4\frac{x^2 + y^2}{w_0^2} - \frac{z^2}{\zR^2} \right\rbrace
\end{equation}
where $w_0$ is the beam waist at the focus and $z_R$ the Rayleigh length.
The half widths of $\psource$ correspond to those of the squared irradiance $I^2$
that typically governs two-photon processes (\latin{c.f.}\ ref.~\citenum{Waller2019}, figure~1).

The dynamics of the surrounding photo-resist is described by the Navier-Stokes equations
\begin{subequations}
    \label{eq:NSE}
    \begin{align}
        \begin{split}
            \rho_0 \pder[\bvec{u}]{t} + \rho_0(\bvec{u} \cdot \nabla)\bvec{u} = {}
            & - \nabla \flpress + \eta \nabla^2 \bvec{u}
            \\
            & + \rho_0 \bvec{g} \big( 1 - \beta (T - \overline T ) \big),
        \end{split}\\
        \nabla\cdot\bvec u = {} & 0
    \end{align}
\end{subequations}
in the Boussinesq approximation\cite{Tritton1988}, where $\bvec{u}$ and $P$ denote fluid velocity and pressure, respectively.
Deviations of the local temperature $T$ from the average $\overline T$
due to the heat that is transferred from particles to fluid constitute the sole driving force.
This buoyancy force is proportional to the fluid's thermal expansion coefficient $\beta$,
the gravitational acceleration $\bvec g$ and the unperturbed fluid density $\rho_0$ at ambient temperature.

Heat propagation within the fluid is described by the heat equation
\begin{equation}
    \label{eq:heat}
    \rho_0 c_p (\dtime T + \bvec u \cdot \nabla T ) = - \nabla\cdot(k\nabla T) + c \crossec{abs} I
\end{equation}
where $k$ and $c_p$ denote the thermal conductivity and the heat capacity, respectively, of the photo-resist.
As the photo-resist is transparent at the laser wavelength,
we consider particles to be the only substance to accumulate heat as they partially
absorb incoming light.
Thus we put the heat source term $c \crossec{abs} I$
as product of local particle concentration $c$, their absorption cross-section $\crossec{abs}$ and irradiance $I$.
This very slightly overestimates the total heat input as it does not account for particles
up the beam axis shading those further down.
Strong shielding, however, would interrupt the laser writing process and thus is ruled out by experience.

%
\subsection{Convection and heating}
\label{sec:convection}
In order to quantify the impact of buoyancy on the MDLW process,
we simulate the influence of
varying laser power on the motion of the photo-resist and the peak temperatures in the vicinity of the focus.

We assume a particle diameter of $45\unit{nm}$
which is a size typically found on SEM images of our samples and
is already large enough to be efficiently trapped and heated.

\begin{figure}
    \includegraphics[width=\columnwidth]{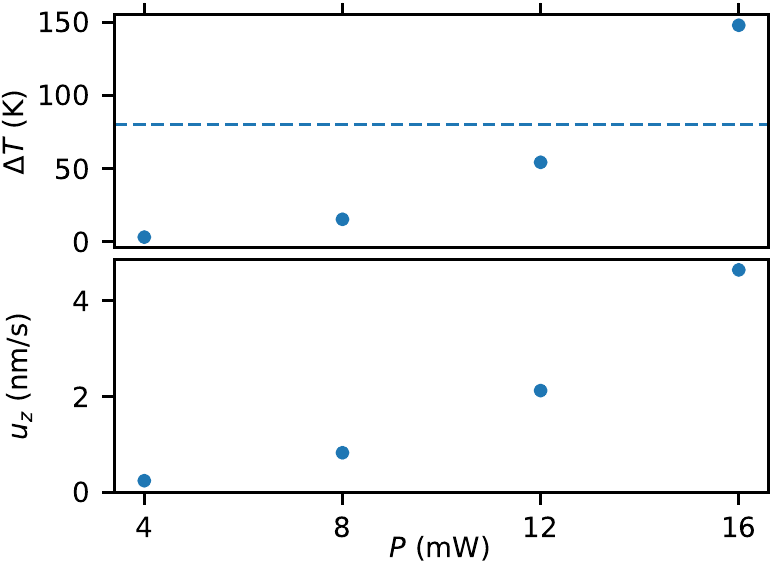}
    \caption{
        Maximum temperature increase $\Delta T$ relative to ambient (top axes)
        and peak fluid velocity $u_z$ (bottom axes) as a function of laser power.
        The dashed line indicates the boiling point assuming $\SI{20}{\degreeCelsius}$ ambient temperature.
        The particle diameter is set to 45\,nm, \textNA=1.2, scan speed $v=10\unit{\micro\m\per\s}$.
        Note that the scaling of both quantities is identical; the fluid velocity behaves proportional to temperature increase.
    }
    \label{fig:deltaTemp_vs_P}
\end{figure}

Depending on the laser power and particle diameter, simulations show that steady state is reached after $50 - 200\unit{\milli \s}$.
We observe a non-linear increase of the peak temperatures with growing laser power $P_0$,
which is attributed to the combined action of increased irradiance and trapping
(\latin{i.e.}\ higher particle concentrations exposed to higher light intensity),
see figure~\ref{fig:deltaTemp_vs_P}.

\begin{figure}
    \includegraphics[width=\columnwidth]{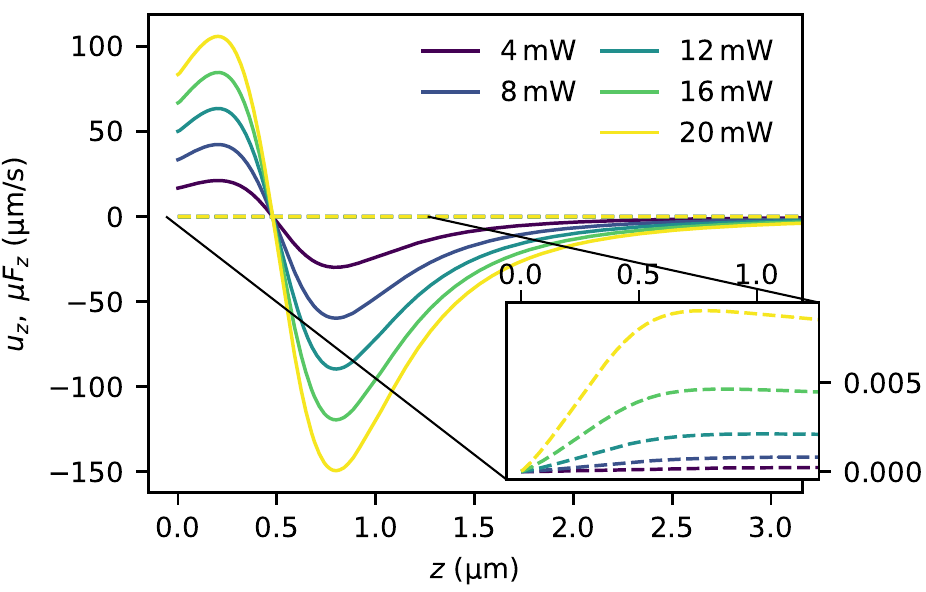}
    \caption{Vertical component of fluid velocity $u_z$ (dashed lines)
        compared to advection of particles $\mob F_z$ due to optical forces (solid lines)
        along beam axis $z$.
        Wavelength 780\,nm, NA=1.2, scan speed $\SI{10}{\micro\m\per\s}$, 45\,nm particles produced at a rate of $\SI{0.52}{\micro \m \cubed \per \s}$.
        Color encodes laser power from 4\,mW (violet) up to 20\,mW (yellow).
        Maximum fluid velocity is reached slightly above focus, see zoomed inset.
    }
    \label{fig:uz_overZ}
\end{figure}

The fluid develops a typical toroidal convection pattern
centered around the beam axis.
Maximum velocity is reached slightly above the focus,
momentum then diffuses laterally which leads to a linear decrease of $u_z$, see figure~\ref{fig:uz_overZ}.

\begin{figure}
    \includegraphics[width=\columnwidth]{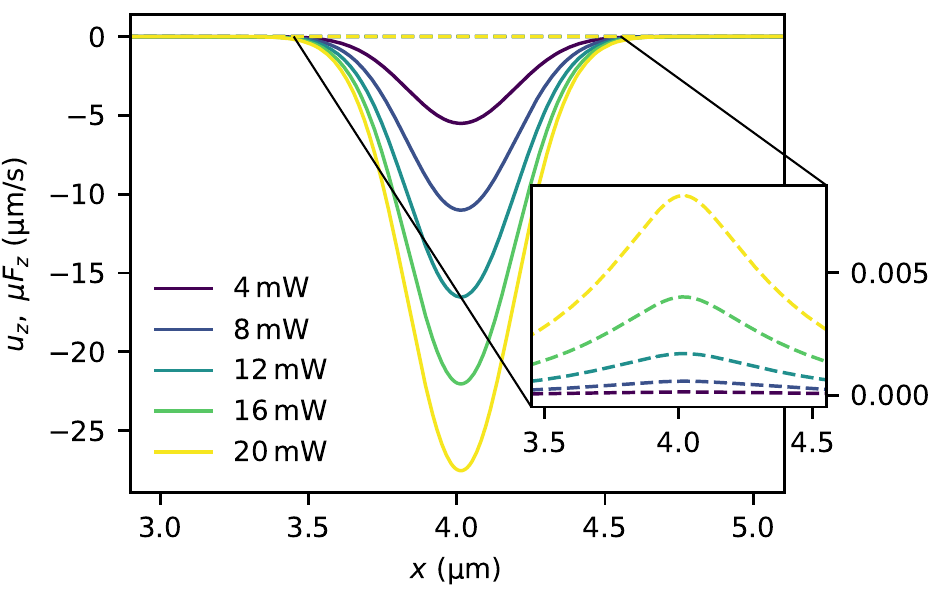}
    \caption{Vertical component of fluid velocity $u_z$ (dashed lines)
        compared to advection of particles $\mob F_z$ due to optical forces (solid lines)
        plotted along a horizontal axis through laser focus.
        Colors and fabrication parameters as given in figure \ref{fig:uz_overZ}.
        Vertical fluid velocity is much smaller than advection due to optical forces,
        see zoomed inset.
    }
    \label{fig:uz_overX}
\end{figure}
As shown in figs.~\ref{fig:uz_overX} and \ref{fig:uz_overZ}, the fluid velocities are negligible
compared to the advection $\mob \Fopt$ of particles by optical forces;
the Peclet numbers which relate diffusion to fluid velocity $\bvec u$ are much smaller than unity, see table \ref{tab:dimlessNumbers}.
\begin{table}[h]
    \caption{Peclet und Reynolds numbers in MDLW}
    \begin{tabularx}{\columnwidth}{@{\extracolsep{\fill}}l l l}
    \toprule
                                    & referring to $\bvec u$              & referring to $\bvec v$         \\
        \midrule
        thermal Peclet  \; $\Pec_T$ & $7\ttenTo{-9}$                   & $ 7\ttenTo{-6}\dots 7\ttenTo{-4}$ \\
        particle Peclet \; $\Pec_c$ & $(0.4 \dots 1.4)\ncdot 10^{-4}$     & $0.1 \dots 10$                  \\
        fluid Reynolds  \; $\Rey$   & $10^{-8}$                           & --                              \\
        \bottomrule
    \end{tabularx}
    \justify
        The Peclet numbers $\Pec_T,\ \Pec_c$ are given with respect to both fluid velocity $\bvec u$ and laser scan speed $\bvec v$;
        only for the latter they approach or surpass unity.
    \label{tab:dimlessNumbers}
\end{table}
Despite particles are permanently produced and spread diffusively from the focus,
they do not present a heat source large enough to cause a problematic amount of buoyancy.
The generality of the maximum convection velocities obtained from our simulations is supported
by two important characteristics inherent to the MDLW process:
First, the temperature is naturally limited by the boiling point.
Second, the spatial extent of the heat source only covers the laser focus,
outside the focal region, the temperature decays reciprocally.
Both limit the total momentum input due to buoyancy into the fluid.

Furthermore, this analysis shows that transport of particles due to the scanning movement of the focus
can be important and impact their local concentrations,
whereas heat diffusion is not expected to be affected by advection, neither of scanning nor fluid motion.

It follows that the back coupling of the fluid velocity $\bvec u$ into eqs.~\eqref{eq:heat},~\eqref{eq:particleDiffusion} is not important,
\latin{i.e.} particle concentration and temperature profile are hardly affected by advection.
Consequently, eqs.~\eqref{eq:heat},~\eqref{eq:particleDiffusion} become linearly scalable in $\psource$
and the advection $\bvec u$ can be substituted by the laser scan speed $\bvec v$.
Regarding the regime of Reynolds numbers, the Navier-Stokes equations are very well approximated by the linear Stokes equations.
Together, these findings allow to generalize the results of a single simulation to different particle production rates $\psource$.

\subsection{Overheating threshold}
\label{sec:explThresh}

\begin{figure}
    \includegraphics[width=\columnwidth]{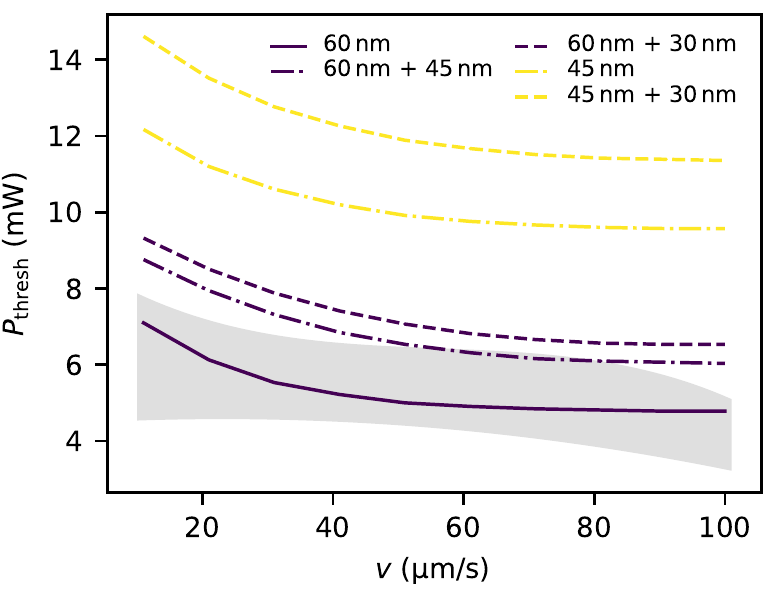}
    \caption{
        Overheating threshold laser powers $\Pthresh$ as a function of scanning speed $v$.
        Grey area indicates data obtained from experiments.
        Simulations were conducted for exemplary particle mixtures with
        either two species in equal parts or a single one.
        Total particle production is adapted to experimentally obtained deposition values.
        Violet lines represent 60\,nm particles (solid) or mixtures in equal parts
        with 45\,nm (dash-dotted) or 30\,nm particles, respectively.
        Yellow lines represent 45\,nm particles (dash-dotted) or a mixture in equal parts
        with 30\,nm particles (dashed).
        Note that a system consisting entirely of 30\,nm particles does not reach the threshold temperature for laser powers up to 20\,mW.
    }
    \label{fig:explThresh}
\end{figure}

As the particle diameter heavily affects the absorption properties and a broad distribution of diameters is present in experiments,
we examine the importance of three exemplary diameters
$d=30, 45, 60\unit{\nano m}$
to the heat generation in a model system with varying fraction of these species.
Smaller diameters are discarded because of their low absorption efficiency;
this assumption is proven by the results of this study, see figure \ref{fig:explThresh}.

Silver structures larger than 60\,nm are rarely observed on SEM-images (see fig.\ \ref{fig:fancyPlot}),
they are mostly found on top of the written lines and at the same time are spread over the entire substrate.
Hence we conclude that they are the result of a post-writing fallout, so they are not immediately produced
during the writing process in the vicinity of the laser focus and thus should not contribute to heat generation.

In simulations, overheating is considered to occur if the temperature $T$
increases to or beyond $\SI{100}{\degreeCelsius}$.
Experimentally, the situation is not as explicit since no clear damage threshold power exists.
Rather, microbubbles increasingly disturb the writing process with increasing laser power.
From microscopy images we therefore define the damage threshold power as the power
at which a written line is no longer connected at at least one position.

Figure \ref{fig:explThresh} displays the experimental threshold region and simulations for a few exemplary particle mixtures;
we consider all cases where a diameter species constitutes either 50\% or 100\% of the total deposited volume.
For laser powers up to $\SI{20}{\milli \watt}$, a system comprising
solely 30\,nm particles does not reach the threshold temperature.
This supports our choice of dismissing particles smaller than 30\,nm from the outset.
The presence of 60\,nm particles significantly lowers the threshold powers
which then match the values obtained from experiments.
Accordingly, the lowest threshold is observed for a system consisting of 60\,nm particles only.
Mixtures containing both 60\,nm particles and one of the smaller diameters in equal shares 
show increased thresholds that differ only very slightly from each other.
Hence, the biggest species present in the system determines the thermal sensitivity.
This is attributed to both the better trapping of bigger particles and the strong increase
of the absorption cross section $\crossec{abs}$ in this size regime.
Together, this results in higher concentrations of particles in the laser focus and a much increased total absorbance.

\subsection{Deposition efficiency}
\label{sec:deposition}

In the following, we examine the influence of optical forces on the deposition of particles
onto the substrate in order to investigate a possible increase of fabrication speeds.
\begin{figure}
    \includegraphics[width=\columnwidth]{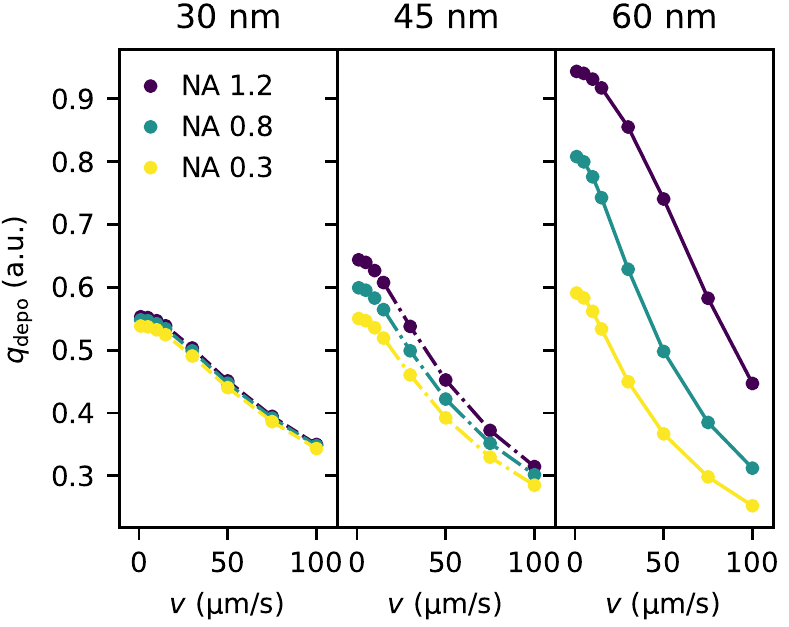}
    \caption{Relative amount of particles deposited at substrate as a function of scan speed.
        Particle diameters from 30\,nm (left) to 60\,nm (right).
        Laser power 6mW, wavelength 780\,nm, \textNA=0.3 (yellow), \textNA=0.8 (green), \textNA=1.2 (violet).
        Towards higher speeds, deposition efficiency decreases stronger for bigger particles; they are flushed out of
        the laser focus where trapping and radiation pressure aid deposition.
        Small particles are less influenced by scan speed since their high diffusivity dominates (small $\Pec$-numbers).
    }
    \label{fig:qdepo_vs_v}
\end{figure}

In simulations, this is carried out by integrating the vertical flux
of particles $\pflux$ onto the substrate, \latin{i.e.} on the bottom face of the simulation domain.
Normalizing this result to the total particle production rate $\psource$
yields the deposition efficiency $\qdepo$ (see methods section for definition and determination thereof)
which, since $\pflux \propto \psource$, is independent of the model input parameter $\psource$.
This dimensionless quantity represents the \textit{physical} efficiency of the MDLW-process
and gives the fraction of particles that adsorb at the structure,
its inverse represents particles that diffuse away and appear as debris on the substrate.
Note that the total efficiency of MDLW comprises the quantum-yield of the 2-photon process,
the chemical reactions induced by the latter and nucleation and re-dissolution of silver.
These processes are not included in the deposition efficiency $\qdepo$ and are generally beyond the scope of this paper.

\begin{figure}
    \includegraphics[width=\columnwidth]{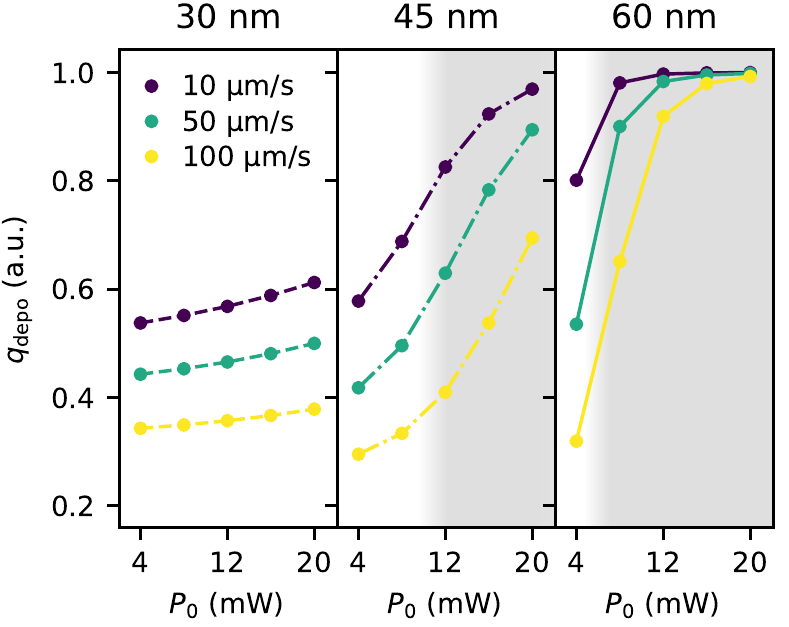}
    \caption{Relative amount of particles deposited at substrate as a function of laser power
        for particle diameters from 30\,nm (left column) to 60\,nm (right column).
        Exemplary scan speeds from $10\unit{\micro\m\per\s}$ (violet) to $100\unit{\micro\m\per\s}$  (yellow),
        wavelength 780\,nm, NA=1.2.
        The gray areas indicate laser powers above the threshold of boiling of the photo-resist obtained
        from simulations for the corresponding particle diameter (see fig.~\ref{fig:explThresh}).
    }
    \label{fig:qdepo_vs_P0}
\end{figure}

Small particles show a low deposition efficiency ($\qdepo(d\!\!=\!\!30\unit{nm})\lesssim 0.5$)
and respond insensitively to variations of optical forces by means of laser power $P_0$ and \textNA, see figs.~\ref{fig:qdepo_vs_v} and \ref{fig:qdepo_vs_P0}.
This is attributed to their higher diffusivity combined with the relatively small susceptibility to optical forces,
leading to small Peclet-numbers and diffusion-dominated behavior (\latin{c.f.}\ table~\ref{tab:dimlessNumbers}).

Larger particles are more affected by changes to the optical setup;
switching from NA=1.2 to NA=0.3 which comes along with much smaller optical force,
halves the deposition efficiency of 60\,nm particles uniformly for all scan speeds
under consideration (\latin{c.f.}\ fig.\ \ref{fig:qdepo_vs_v}, right graph).
On the contrary, increasing the laser power and a high \textNA\ greatly improve deposition efficiency.
It approaches unity for 60\,nm particles at moderate scan speeds up to $\approx\! 30\unit{\micro\m\per\s}$,
see center and right graph of figure~\ref{fig:qdepo_vs_P0}.
At higher scan speeds (\latin{c.f.}\ fig.\ \ref{fig:qdepo_vs_v}),
we observe a significant decrease of $\qdepo$ for all particle species.
This effect could be mitigated by increased laser power,
while in the bounds for $P_0$ given above.
Therefore we conclude that trap strength is important for the efficient deposition of particles in MDLW.

AFM height maps of silver lines fabricated at different scan speeds provide the cross-sections $A$ and the surface roughness of these lines (not shown here).
Multiplication with the scan speed yields the net volumetric silver production rate $S_V = A v$.
For a given diameter, the numeric particle production rate $\psource$,
which is fed into eq. \eqref{eq:particleDiffusion}, follows from division
by the particle volume corresponding to that diameter.
\begin{table}
    \caption{Analysis of AFM measured cross-sections}
    \begin{tabularx}{\columnwidth}{XXX}
        \toprule
                   & $g\ \si{(\micro\m^3\per\s)}$  & $b\ \si{(\micro\m^2)}$ \\
        \midrule
        $\NA\,0.3$      & 0.401(30)  & 0.0786(31) \\
        $\NA\,0.8$      & 0.212(19)  & 0.0548(14) \\
        $\NA\,1.2$      & 0.0687(54) & 0.0263(7)  \\
        $\NA\,1.2$, \#2 & 0.264(83)  & 0.0187(20) \\
        \bottomrule
    \end{tabularx}
    \justify
        Parameters obtained from error-weighted least-square fits of eq. \eqref{eq:reciFit} to line cross-sections obtained from AFM measurements.
        Scan speed ranges from $3\dots21\unit{\micro\m\per\s}$ and laser power was 3\,mW.
        The second run with \textNA=1.2 was produced with a different batch of photo-resist and with scan speeds from $11\dots101\unit{\micro\m\per\s}$.
    \label{tab:fitParams}
\end{table}

The line cross-sections show a behavior similar to the expected reciprocal decay with increasing scan speed.
Thus we fitted line cross-sections $A$ with
\begin{equation}
      A(v) = \frac{g}{v} + b
    \label{eq:reciFit}
\end{equation}
The parameters $g$ and $b$ are subject to optimization by the fit routine,
see table \ref{tab:fitParams} for results.

We compared the residuals of fits of eq. \eqref{eq:reciFit} and the candidate modifications
\begin{subequations}
    \begin{gather}
        \label{eq:fitQdepoNoBaseline}
        A(v) = \qdepo(d, v) \frac{g}{v} + b, \\
        \label{eq:fitQdepoBaseline}
        A(v) =\qdepo(d, v) \left( \frac{g}{v} + b \right).
    \end{gather}
\end{subequations}
These modifications yield the parameters $g,b$ corresponding to the gross particle production in consideration of the associated theoretical deposition efficiency.
Fits of eq.\ \eqref{eq:fitQdepoNoBaseline} describe experimental data very well
and yield residuals practically identical to those obtained from the unmodified fit function.
Eq.\ \eqref{eq:fitQdepoBaseline} fails to describe experimental data at scan speeds above $\SI{60}{\micro \m\per \s}$.
Hence, the baseline production $b$ and associated deposition are insensitive to the scan speed;
we suggest that the baseline-production is promoted by the substrate and these particles aggregate immediately.

\section{Conclusions}
We introduced a model for the simulation of direct laser writing of metallic structures (MDLW)
and discussed its results concerning convection, heating and deposition.

Convection of photo-resist does not affect the deposition of particles.
Since rapidly increasing velocity with growing heat source has been reported \cite{Donner2011},
care must be taken, if the heated, \latin{i.e.}\ illuminated, area is increased,
\latin{e.g.}, if a second laser is added for enhanced trapping.
Furthermore, other effects such as thermophoresis and surface tension gradient driven (Marangoni) flows should be investigated in future to further refine the model.

The comparison of maximum feasible laser powers obtained from simulations and experiments, respectively,
suggested that to disturb the MDLW process critically, the onset of evaporation is already sufficient, before permanent boiling occurs.
The impact of clusters of particles which potentially promote
heat absorption could be investigated using molecular dynamics
simulations which resolve individual particles and solving for the electric fields at the same time.

Our simulations suggest that the deposition of particles improves with increasing trapping strength.
This could also mitigate losses due to increased scan speed, especially for large particles.

AFM measurements of silver structures produced via MDLW delivered the required input parameter for the model equations,
as well as the surface roughness (not shown here).
The increase of the latter towards slower scan speeds indicates
that the growth of larger particles then is promoted.
Hence a better understanding of growth mechanisms and time-scales is desirable.

\section{Methods}
\small
\subsection{Model parameters}

For the investigation of convection, we set the scan speed $v_x = \SI{10}{\micro \m \per \s}$.
Unless indicated otherwise, we assume an effective numerical aperture of \textNA=1.2
which accounts for the lowered refractive index of the photo resist compared to immersion oil.

For the refractive index of silver particles, we use the values published by Johnson and
Christy\cite{JC72} for the corresponding wavelength.
Although originally measured for thin films, this is proven practice\cite{Billaud2007}.
We treat particle sizes ranging from $\dpart = 30 \dots 60\unit{nm}$ in our simulations,
this range is deduced from inspection of SEM-images.
Particles are generally assumed to be spherical.

The remaining parameters appearing in eqs.\ \eqref{eq:particleDiffusion}, \eqref{eq:heat} and \eqref{eq:NSE}
are set to the values given in table \ref{tab:modelParameters}.
\begin{table}[h!]
    \caption{Model parameters used in eqs.\ \eqref{eq:gradForce}, \eqref{eq:NSE}, \eqref{eq:heat}}
    \begin{tabularx}{\columnwidth}{@{\extracolsep{\fill}}l l l}
        \toprule
        & Symbol  & Value \\
        \midrule
        kinematic viscosity     & $\eta$   & $0.001\,\si{\Pa\s}$     \\
        fluid density           & $\rho_0$ & $1000\,\si{\kg\per\m^3}$     \\
        fluid heat capacity     & $c_p$    & $4184\,\si{\J\per(\kg \K)}$    \\
        fluid heat expansion coefficient   & $\beta$ & $0.002\unit{\per\K}$ \\
        fluid heat conductivity & $k$      & $0.6 \,\si{W/(K\, m)}$  \\
        ambient temperature     & $T_0$    & $293\,\si{K}$           \\
        refractive index        & \nliq    & 1.33                    \\
        \bottomrule
    \end{tabularx}
    \label{tab:modelParameters}
\end{table}

\subsection{Details of simulation model}
\label{ap:numerics}
The model equations
\eqref{eq:particleDiffusion}, \eqref{eq:NSE} and \eqref{eq:heat}
are solved within a cubic domain $\Omega$ of $8\,\si{\micro\m}$
edge length, centered around the beam axis.
This domain is much larger than the focus itself which has proven necessary so that the system
is not perturbed by the influence of the boundary $\Gamma \equiv \partial \Omega$.
The simulation domain axially centered around the laser beam
and the focus is $0.5\,\si{\micro\m}$ above the domain's bottom face.
The latter represents a substrate which is being written on.

We apply homogeneous Dirichlet boundary conditions $c\big|_\Gamma = 0$
to particle diffusion \eqref{eq:particleDiffusion} on all walls, which is also known as absorbing boundaries.
This is due to the fact that particles which meet the substrate stick there and do not go back into suspension.
Also, particles that depart far enough from the focal region are lost and either re-dissolve or appear as debris elsewhere on the sample.

For the fluid dynamics \eqref{eq:NSE},
we consider a reference frame that moves with the constant scan speed $\bvec v$ of the laser.
Consequently, the initial fluid velocity is set to that of the laser:
\begin{equation}
    \bvec u\big|_{t=0} = \bvec v
\end{equation}
Correspondingly, we assume the substrate, that is the bottom face $\Gamma^-$ at $z=0$,
to correspond to a no-slip boundary that moves with $-\bvec v$:
\begin{align}
    \bvec{u}\big|_{\Gamma^-} = -\bvec v, \quad
    \pder[\flpress]{\bvec n}\Big|_{\Gamma^-} = 0.
\end{align}
At the same time, side and top walls $\Gamma^\Box$, $\Gamma^+$, respectively,
shall be as ``transparent'' as possible for the fluid,
which we achieve by applying homogeneous Neumann boundary conditions both for pressure and velocity:
\begin{equation}
    (\nabla\bvec u) \cdot \bvec n \big|_{\Gamma^\Box, \Gamma^+} = 0, \quad
    \pder[\flpress]{\bvec n}\Big|_{\Gamma^\Box} = 0.
\end{equation}
Thereby we do not constrain the size of convection eddies.
The reference pressure is set at the top face $\Gamma^+$ at $z=8\,\si{\micro\m}$ via a homogeneous Dirichlet condition:
\begin{align}
    \flpress\big|_{\Gamma^+} = 0.
\end{align}

We assume the substrate to function as heat sink and,
in order to avoid an unphysical accumulation of heat within the system,
fix the boundaries
\begin{equation}
    T\big|_\Gamma = T_0
\end{equation}
to the ambient temperature for the heat equation~\eqref{eq:heat}.
Due to the sufficient size of our simulation domain,
this hardly affects temperatures at the laser focus.

The set of coupled equations \eqref{eq:particleDiffusion}, \eqref{eq:NSE}, \eqref{eq:heat}
is solved numerically using the multi-purpose finite volume solver \textit{Corheos}\cite{Corheos2010, Corheos2013}
which applies fully implicit Euler time-stepping.

\subsection{Laser beam}
\label{ap:beamFormulas}
Laser beams are modeled as paraxial Gaussian beam\cite{Novotny2009}.
For the laser's lateral beam waist $w_0$ and Rayleigh length $\zR$ at a given numerical aperture $\NA$,
we assume
\begin{align}
    w_0 = \frac{1.22 \lam}{2 \NA \sqrt{(2 \ln 2)} }, \\
    \zR = \frac{\lam}{2\Big( \nliq^2 - \sqrt{\nliq^2 - \NA^2} \Big)},
\end{align}
respectively.
Albeit the laser emits short pulses, we use the forces of the averaged laser power
since pulsed lasers show at least equal trapping efficiencies as the equivalent CW laser\cite{Agate04, Shane2010}.

\subsection{Overheating threshold}
\label{ap:superposition}
In order to obtain the threshold laser power $\Pthresh$ from the discretely spaced simulation results $T(P_0, d, \bvec v)$,
we interpolate these results with respect to $P_0$ and $v$ by means of a two dimensional cubic spline.
For a given particle diameter, this immediately yields $\Pthresh$ as a function of scan speed.

The threshold laser power for a mixture $s$ of particle species is then obtained by
exploiting the linearity of eqs.~\eqref{eq:particleDiffusion}, \eqref{eq:heat}
in the particle source term $\psource$ as established above.
A posteriori, we superpose simulations of particle diameters $d = 30, 45, 60\unit{nm}$, respectively,
such that the total volumetric silver production remains fixed at the value
obtained from experimentally measured line cross-sections.
In order to maintain consistency with the latter, the particle production rates
must be corrected for the fraction of particles $\qdepo(P_0,d,\bvec v)$
that effectively reach the substrate.
Then, for the gross particle production $\psourced$ of each diameter species $d$, we write
\begin{equation}
    \psourced(P_0, \bvec v) \equiv \frac{\psource \, \psourcefrac}{q(d, P_0, \bvec v)}
\end{equation}
where $\psourcefrac$ represents the net fraction
of deposited particles with diameter $d$ within the system and the relation
\begin{equation}
    \label{eq:simplex}
    \sum_d \psourcefrac = 1
\end{equation}
holds.
Such a superposition yields the temperature $T(s, P_0, \bvec v)$ for all laser powers $P_0$
and scan speeds $v$ considered in the parameter space.
We obtain the threshold powers by means of a greedy search for $P_0$ s.t. $T(s, P_0, \bvec v) = \SI{100}{\degreeCelsius}$.

For determination of the dependence of temperature on scan speed and laser power,
simulations were run with $P_0 =  4\dots 20\unit{\milli\W}$,
while the scan speed was sampled non-uniformly at $v_x = 1, 5, 10, 15, 30, 50, 75, 100\unit{\micro\m\per\s}$
in order to capture all features at lower velocities.

\subsection{Deposition efficiency}
\label{ap:qdepo}
In order to measure the fraction of particles that adsorb to the line being written,
we integrate the vertical flux of particles
\begin{equation}
    \label{eq:flux}
    \pflux_z =  (\mu \Fopt\, c - D \nabla c)_z
\end{equation}
on the substrate (i.e. on the bottom face of the domain)
over a square of $2\,\si{\micro\m}$ edge length, centered around the beam axis.
Normalizing this result to the total particle production rate $\psource$ yields
\begin{equation}
    \qdepo = \frac{1}{\psource}\int_{\fluxIntBound} \pflux_z \dd A,
\end{equation}
which, since $\pflux \propto \psource$, is independent of the model input parameter $\psource$.

\subsection{Structure fabrication}
Silver lines are fabricated using a commercially available direct laser writing system (Nanoscribe).
This system uses a tightly focussed laser beam (80 MHz, 780 nm) to locally initiate photo-reactions.
In our experiment, a liquid photosensitive material consisting of 0.4 M silver perchlorate, 0.05 M trisodium citrate,
20 v\% ammonia water and water, is sandwiched between two BK 7 glass substrates using \SI{50}{\micro\m} tape spacers.
Structures are fabricated on the bottom facet of the top glass substrates.
Thus, radiation pressure points towards the substrates.

\subsection{AFM measurements and evaluation}
AFM height maps of the fabricated silver lines are taken using a JPK Nanowizard~3 AFM.
Eleven parallel lines are fabricated with a separation of \SI{2}{\micro\m}.
A Matlab script is used to automatically derive the averaged line widths,
line heights and RMS roughnesses from these height maps.

\normalsize

\bibliography{MDLW.bib}

\providecommand{\latin}[1]{#1}
\providecommand*\mcitethebibliography{\thebibliography}
\csname @ifundefined\endcsname{endmcitethebibliography}
  {\let\endmcitethebibliography\endthebibliography}{}
\begin{mcitethebibliography}{41}
\providecommand*\natexlab[1]{#1}
\providecommand*\mciteSetBstSublistMode[1]{}
\providecommand*\mciteSetBstMaxWidthForm[2]{}
\providecommand*\mciteBstWouldAddEndPuncttrue
  {\def\EndOfBibitem{\unskip.}}
\providecommand*\mciteBstWouldAddEndPunctfalse
  {\let\EndOfBibitem\relax}
\providecommand*\mciteSetBstMidEndSepPunct[3]{}
\providecommand*\mciteSetBstSublistLabelBeginEnd[3]{}
\providecommand*\EndOfBibitem{}
\mciteSetBstSublistMode{f}
\mciteSetBstMaxWidthForm{subitem}{(\alph{mcitesubitemcount})}
\mciteSetBstSublistLabelBeginEnd
  {\mcitemaxwidthsubitemform\space}
  {\relax}
  {\relax}

\bibitem[Waller \latin{et~al.}(2021)Waller, Karst, and von
  Freymann]{Waller2021}
Waller,~E.~H.; Karst,~J.; von Freymann,~G. \emph{Light: Advanced Manufacturing}
  \textbf{2021}, \emph{2}, 1\relax
\mciteBstWouldAddEndPuncttrue
\mciteSetBstMidEndSepPunct{\mcitedefaultmidpunct}
{\mcitedefaultendpunct}{\mcitedefaultseppunct}\relax
\EndOfBibitem
\bibitem[Thiele \latin{et~al.}(2017)Thiele, Arzenbacher, Gissibl, Giessen, and
  Herkommer]{Thiele2017}
Thiele,~S.; Arzenbacher,~K.; Gissibl,~T.; Giessen,~H.; Herkommer,~A.~M.
  \emph{Science Advances} \textbf{2017}, \emph{3}, e1602655\relax
\mciteBstWouldAddEndPuncttrue
\mciteSetBstMidEndSepPunct{\mcitedefaultmidpunct}
{\mcitedefaultendpunct}{\mcitedefaultseppunct}\relax
\EndOfBibitem
\bibitem[Thiele \latin{et~al.}(2019)Thiele, Pruss, Herkommer, and
  Giessen]{Thiele2019}
Thiele,~S.; Pruss,~C.; Herkommer,~A.~M.; Giessen,~H. \emph{Optics Express}
  \textbf{2019}, \emph{27}, 35621\relax
\mciteBstWouldAddEndPuncttrue
\mciteSetBstMidEndSepPunct{\mcitedefaultmidpunct}
{\mcitedefaultendpunct}{\mcitedefaultseppunct}\relax
\EndOfBibitem
\bibitem[Toulouse \latin{et~al.}(2021)Toulouse, Drozella, Thiele, Giessen, and
  Herkommer]{Toulouse2021}
Toulouse,~A.; Drozella,~J.; Thiele,~S.; Giessen,~H.; Herkommer,~A. \emph{Light:
  Advanced Manufacturing} \textbf{2021}, \emph{2}, 1--11\relax
\mciteBstWouldAddEndPuncttrue
\mciteSetBstMidEndSepPunct{\mcitedefaultmidpunct}
{\mcitedefaultendpunct}{\mcitedefaultseppunct}\relax
\EndOfBibitem
\bibitem[Schumann \latin{et~al.}(2014)Schumann, Bückmann, Gruhler, Wegener,
  and Pernice]{Schumann2014}
Schumann,~M.; Bückmann,~T.; Gruhler,~N.; Wegener,~M.; Pernice,~W. \emph{Light:
  Science {\&} Applications} \textbf{2014}, \emph{3}, e175--e175\relax
\mciteBstWouldAddEndPuncttrue
\mciteSetBstMidEndSepPunct{\mcitedefaultmidpunct}
{\mcitedefaultendpunct}{\mcitedefaultseppunct}\relax
\EndOfBibitem
\bibitem[Deubel \latin{et~al.}(2004)Deubel, von Freymann, Wegener, Pereira,
  Busch, and Soukoulis]{Deubel2004}
Deubel,~M.; von Freymann,~G.; Wegener,~M.; Pereira,~S.; Busch,~K.;
  Soukoulis,~C.~M. \emph{Nature Materials} \textbf{2004}, \emph{3},
  444--447\relax
\mciteBstWouldAddEndPuncttrue
\mciteSetBstMidEndSepPunct{\mcitedefaultmidpunct}
{\mcitedefaultendpunct}{\mcitedefaultseppunct}\relax
\EndOfBibitem
\bibitem[Rill \latin{et~al.}(2008)Rill, Plet, Thiel, Staude, von Freymann,
  Linden, and Wegener]{Rill2008}
Rill,~M.~S.; Plet,~C.; Thiel,~M.; Staude,~I.; von Freymann,~G.; Linden,~S.;
  Wegener,~M. \emph{Nature Materials} \textbf{2008}, \emph{7}, 543--546\relax
\mciteBstWouldAddEndPuncttrue
\mciteSetBstMidEndSepPunct{\mcitedefaultmidpunct}
{\mcitedefaultendpunct}{\mcitedefaultseppunct}\relax
\EndOfBibitem
\bibitem[Gansel \latin{et~al.}(2009)Gansel, Thiel, Rill, Decker, Bade, Saile,
  von Freymann, Linden, and Wegener]{Gansel2009}
Gansel,~J.~K.; Thiel,~M.; Rill,~M.~S.; Decker,~M.; Bade,~K.; Saile,~V.; von
  Freymann,~G.; Linden,~S.; Wegener,~M. \emph{Science} \textbf{2009},
  \emph{325}, 1513--1515\relax
\mciteBstWouldAddEndPuncttrue
\mciteSetBstMidEndSepPunct{\mcitedefaultmidpunct}
{\mcitedefaultendpunct}{\mcitedefaultseppunct}\relax
\EndOfBibitem
\bibitem[Gansel \latin{et~al.}(2012)Gansel, Latzel, Frölich, Kaschke, Thiel,
  and Wegener]{Gansel2012}
Gansel,~J.~K.; Latzel,~M.; Frölich,~A.; Kaschke,~J.; Thiel,~M.; Wegener,~M.
  \emph{Applied Physics Letters} \textbf{2012}, \emph{100}, 101109\relax
\mciteBstWouldAddEndPuncttrue
\mciteSetBstMidEndSepPunct{\mcitedefaultmidpunct}
{\mcitedefaultendpunct}{\mcitedefaultseppunct}\relax
\EndOfBibitem
\bibitem[Tanaka \latin{et~al.}(2006)Tanaka, Ishikawa, and Kawata]{Tanaka2006}
Tanaka,~T.; Ishikawa,~A.; Kawata,~S. \emph{Applied Physics Letters}
  \textbf{2006}, \emph{88}, 081107\relax
\mciteBstWouldAddEndPuncttrue
\mciteSetBstMidEndSepPunct{\mcitedefaultmidpunct}
{\mcitedefaultendpunct}{\mcitedefaultseppunct}\relax
\EndOfBibitem
\bibitem[Blasco \latin{et~al.}(2016)Blasco, Müller, Müller, Trouillet,
  Schön, Scherer, Barner-Kowollik, and Wegener]{Blasc2016}
Blasco,~E.; Müller,~J.; Müller,~P.; Trouillet,~V.; Schön,~M.; Scherer,~T.;
  Barner-Kowollik,~C.; Wegener,~M. \emph{Advanced Materials} \textbf{2016},
  \emph{28}, 3592--3595\relax
\mciteBstWouldAddEndPuncttrue
\mciteSetBstMidEndSepPunct{\mcitedefaultmidpunct}
{\mcitedefaultendpunct}{\mcitedefaultseppunct}\relax
\EndOfBibitem
\bibitem[Greer and Street(2007)Greer, and Street]{Greer2007}
Greer,~J.~R.; Street,~R.~A. \emph{Acta Materialia} \textbf{2007}, \emph{55},
  6345--6349\relax
\mciteBstWouldAddEndPuncttrue
\mciteSetBstMidEndSepPunct{\mcitedefaultmidpunct}
{\mcitedefaultendpunct}{\mcitedefaultseppunct}\relax
\EndOfBibitem
\bibitem[Vyatskikh \latin{et~al.}(2018)Vyatskikh, Delalande, Kudo, Zhang,
  Portela, and Greer]{Greer2018}
Vyatskikh,~A.; Delalande,~S.; Kudo,~A.; Zhang,~X.; Portela,~C.~M.; Greer,~J.~R.
  \emph{Nature Communications} \textbf{2018}, \emph{9}\relax
\mciteBstWouldAddEndPuncttrue
\mciteSetBstMidEndSepPunct{\mcitedefaultmidpunct}
{\mcitedefaultendpunct}{\mcitedefaultseppunct}\relax
\EndOfBibitem
\bibitem[Waller \latin{et~al.}(2019)Waller, Dix, Gutsche, Widera, and von
  Freymann]{Waller2019}
Waller,~E.~H.; Dix,~S.; Gutsche,~J.; Widera,~A.; von Freymann,~G.
  \emph{Micromachines} \textbf{2019}, \emph{10}, 827\relax
\mciteBstWouldAddEndPuncttrue
\mciteSetBstMidEndSepPunct{\mcitedefaultmidpunct}
{\mcitedefaultendpunct}{\mcitedefaultseppunct}\relax
\EndOfBibitem
\bibitem[Waller and von Freymann(2018)Waller, and von Freymann]{Waller2018}
Waller,~E.~H.; von Freymann,~G. \emph{Nanophotonics} \textbf{2018}, \emph{7},
  1259--1277\relax
\mciteBstWouldAddEndPuncttrue
\mciteSetBstMidEndSepPunct{\mcitedefaultmidpunct}
{\mcitedefaultendpunct}{\mcitedefaultseppunct}\relax
\EndOfBibitem
\bibitem[Lee \latin{et~al.}(2017)Lee, Lee, Yang, Koh, Lay, Lee, Phang, and
  Ling]{Lee2017}
Lee,~M.~R.; Lee,~H.~K.; Yang,~Y.; Koh,~C. S.~L.; Lay,~C.~L.; Lee,~Y.~H.;
  Phang,~I.~Y.; Ling,~X.~Y. \emph{{ACS} Applied Materials {\&} Interfaces}
  \textbf{2017}, \emph{9}, 39584--39593\relax
\mciteBstWouldAddEndPuncttrue
\mciteSetBstMidEndSepPunct{\mcitedefaultmidpunct}
{\mcitedefaultendpunct}{\mcitedefaultseppunct}\relax
\EndOfBibitem
\bibitem[He \latin{et~al.}(2017)He, Zheng, Dong, Liu, Duan, and Zhao]{He2017c}
He,~G.-C.; Zheng,~M.-L.; Dong,~X.-Z.; Liu,~J.; Duan,~X.-M.; Zhao,~Z.-S.
  \emph{{AIP} Advances} \textbf{2017}, \emph{7}, 035203\relax
\mciteBstWouldAddEndPuncttrue
\mciteSetBstMidEndSepPunct{\mcitedefaultmidpunct}
{\mcitedefaultendpunct}{\mcitedefaultseppunct}\relax
\EndOfBibitem
\bibitem[Baffou \latin{et~al.}(2009)Baffou, Quidant, and Girard]{Baffo2009}
Baffou,~G.; Quidant,~R.; Girard,~C. \emph{Applied Physics Letters}
  \textbf{2009}, \emph{94}, 153109\relax
\mciteBstWouldAddEndPuncttrue
\mciteSetBstMidEndSepPunct{\mcitedefaultmidpunct}
{\mcitedefaultendpunct}{\mcitedefaultseppunct}\relax
\EndOfBibitem
\bibitem[Baffou \latin{et~al.}(2010)Baffou, Quidant, and de~Abajo]{Baffo2010}
Baffou,~G.; Quidant,~R.; de~Abajo,~F. J.~G. \emph{{ACS} Nano} \textbf{2010},
  \emph{4}, 709--716\relax
\mciteBstWouldAddEndPuncttrue
\mciteSetBstMidEndSepPunct{\mcitedefaultmidpunct}
{\mcitedefaultendpunct}{\mcitedefaultseppunct}\relax
\EndOfBibitem
\bibitem[Baffou and Rigneault(2011)Baffou, and Rigneault]{Baffo2011}
Baffou,~G.; Rigneault,~H. \emph{Physical Review B} \textbf{2011},
  \emph{84}\relax
\mciteBstWouldAddEndPuncttrue
\mciteSetBstMidEndSepPunct{\mcitedefaultmidpunct}
{\mcitedefaultendpunct}{\mcitedefaultseppunct}\relax
\EndOfBibitem
\bibitem[Baffou \latin{et~al.}(2013)Baffou, Berto, Ure{\~{n}}a, Quidant,
  Monneret, Polleux, and Rigneault]{Baffo2013}
Baffou,~G.; Berto,~P.; Ure{\~{n}}a,~E.~B.; Quidant,~R.; Monneret,~S.;
  Polleux,~J.; Rigneault,~H. \emph{{ACS} Nano} \textbf{2013}, \emph{7},
  6478--6488\relax
\mciteBstWouldAddEndPuncttrue
\mciteSetBstMidEndSepPunct{\mcitedefaultmidpunct}
{\mcitedefaultendpunct}{\mcitedefaultseppunct}\relax
\EndOfBibitem
\bibitem[Donner \latin{et~al.}(2011)Donner, Baffou, McCloskey, and
  Quidant]{Donner2011}
Donner,~J.~S.; Baffou,~G.; McCloskey,~D.; Quidant,~R. \emph{{ACS} Nano}
  \textbf{2011}, \emph{5}, 5457--5462\relax
\mciteBstWouldAddEndPuncttrue
\mciteSetBstMidEndSepPunct{\mcitedefaultmidpunct}
{\mcitedefaultendpunct}{\mcitedefaultseppunct}\relax
\EndOfBibitem
\bibitem[Ashkin(0000)]{Ashkin_1992}
Ashkin,~A. \emph{Biophysical Journal} \textbf{0000}, \emph{61}, 569--582\relax
\mciteBstWouldAddEndPuncttrue
\mciteSetBstMidEndSepPunct{\mcitedefaultmidpunct}
{\mcitedefaultendpunct}{\mcitedefaultseppunct}\relax
\EndOfBibitem
\bibitem[Ashkin(2000)]{Ashkin_2000}
Ashkin,~A. \emph{{IEEE} Journal of Selected Topics in Quantum Electronics}
  \textbf{2000}, \emph{6}\relax
\mciteBstWouldAddEndPuncttrue
\mciteSetBstMidEndSepPunct{\mcitedefaultmidpunct}
{\mcitedefaultendpunct}{\mcitedefaultseppunct}\relax
\EndOfBibitem
\bibitem[Svoboda and Block(1994)Svoboda, and Block]{Svobo1994}
Svoboda,~K.; Block,~S.~M. \emph{Opt. Lett.} \textbf{1994}, \emph{19},
  930--932\relax
\mciteBstWouldAddEndPuncttrue
\mciteSetBstMidEndSepPunct{\mcitedefaultmidpunct}
{\mcitedefaultendpunct}{\mcitedefaultseppunct}\relax
\EndOfBibitem
\bibitem[Nieminen \latin{et~al.}(2007)Nieminen, Loke, Stilgoe, Knöner,
  Bra{\'{n}}czyk, Heckenberg, and Rubinsztein-Dunlop]{Nieminen2007}
Nieminen,~T.~A.; Loke,~V. L.~Y.; Stilgoe,~A.~B.; Knöner,~G.;
  Bra{\'{n}}czyk,~A.~M.; Heckenberg,~N.~R.; Rubinsztein-Dunlop,~H.
  \emph{Journal of Optics A: Pure and Applied Optics} \textbf{2007}, \emph{9},
  S196--S203\relax
\mciteBstWouldAddEndPuncttrue
\mciteSetBstMidEndSepPunct{\mcitedefaultmidpunct}
{\mcitedefaultendpunct}{\mcitedefaultseppunct}\relax
\EndOfBibitem
\bibitem[Nieminen \latin{et~al.}(2014)Nieminen, du~Preez-Wilkinson, Stilgoe,
  Loke, Bui, and Rubinsztein-Dunlop]{Nieminen2014}
Nieminen,~T.~A.; du~Preez-Wilkinson,~N.; Stilgoe,~A.~B.; Loke,~V.~L.;
  Bui,~A.~A.; Rubinsztein-Dunlop,~H. \emph{Journal of Quantitative Spectroscopy
  and Radiative Transfer} \textbf{2014}, \emph{146}, 59--80\relax
\mciteBstWouldAddEndPuncttrue
\mciteSetBstMidEndSepPunct{\mcitedefaultmidpunct}
{\mcitedefaultendpunct}{\mcitedefaultseppunct}\relax
\EndOfBibitem
\bibitem[Melzer and McLeod(2018)Melzer, and McLeod]{Melzer2018}
Melzer,~J.~E.; McLeod,~E. \emph{{ACS} Nano} \textbf{2018}, \emph{12},
  2440--2447\relax
\mciteBstWouldAddEndPuncttrue
\mciteSetBstMidEndSepPunct{\mcitedefaultmidpunct}
{\mcitedefaultendpunct}{\mcitedefaultseppunct}\relax
\EndOfBibitem
\bibitem[Baffou \latin{et~al.}(2010)Baffou, Quidant, and Girard]{Baffou2010}
Baffou,~G.; Quidant,~R.; Girard,~C. \emph{Physical Review B} \textbf{2010},
  \emph{82}\relax
\mciteBstWouldAddEndPuncttrue
\mciteSetBstMidEndSepPunct{\mcitedefaultmidpunct}
{\mcitedefaultendpunct}{\mcitedefaultseppunct}\relax
\EndOfBibitem
\bibitem[Skylar-Scott \latin{et~al.}(2016)Skylar-Scott, Gunasekaran, and
  Lewis]{Skylar2016}
Skylar-Scott,~M.~A.; Gunasekaran,~S.; Lewis,~J.~A. \emph{Proceedings of the
  National Academy of Sciences} \textbf{2016}, \emph{113}, 6137--6142\relax
\mciteBstWouldAddEndPuncttrue
\mciteSetBstMidEndSepPunct{\mcitedefaultmidpunct}
{\mcitedefaultendpunct}{\mcitedefaultseppunct}\relax
\EndOfBibitem
\bibitem[Bohren and Huffman(1983)Bohren, and Huffman]{BohrenHuffman1983}
Bohren,~C.~F.; Huffman,~D. \emph{Absorption and scattering of light by small
  particles}; Wiley, 1983\relax
\mciteBstWouldAddEndPuncttrue
\mciteSetBstMidEndSepPunct{\mcitedefaultmidpunct}
{\mcitedefaultendpunct}{\mcitedefaultseppunct}\relax
\EndOfBibitem
\bibitem[Gillespie and Seitaridou(2012)Gillespie, and
  Seitaridou]{Gillespie2012}
Gillespie,~D.~T.; Seitaridou,~E. \emph{Simple Brownian Diffusion}; Oxford
  University Press, 2012\relax
\mciteBstWouldAddEndPuncttrue
\mciteSetBstMidEndSepPunct{\mcitedefaultmidpunct}
{\mcitedefaultendpunct}{\mcitedefaultseppunct}\relax
\EndOfBibitem
\bibitem[Tritton(1988)]{Tritton1988}
Tritton,~D. \emph{Physical Fluid Dynamics}; Oxford Science Publ; Clarendon
  Press, 1988\relax
\mciteBstWouldAddEndPuncttrue
\mciteSetBstMidEndSepPunct{\mcitedefaultmidpunct}
{\mcitedefaultendpunct}{\mcitedefaultseppunct}\relax
\EndOfBibitem
\bibitem[Johnson and Christy(1972)Johnson, and Christy]{JC72}
Johnson,~P.~B.; Christy,~R.~W. \emph{Physical Review B} \textbf{1972},
  \emph{6}, 4370--4379\relax
\mciteBstWouldAddEndPuncttrue
\mciteSetBstMidEndSepPunct{\mcitedefaultmidpunct}
{\mcitedefaultendpunct}{\mcitedefaultseppunct}\relax
\EndOfBibitem
\bibitem[Billaud \latin{et~al.}(2007)Billaud, Huntzinger, Cottancin,
  Lerm{\'{e}}, Pellarin, Arnaud, Broyer, Fatti, and Vall{\'{e}}e]{Billaud2007}
Billaud,~P.; Huntzinger,~J.-R.; Cottancin,~E.; Lerm{\'{e}},~J.; Pellarin,~M.;
  Arnaud,~L.; Broyer,~M.; Fatti,~N.~D.; Vall{\'{e}}e,~F. \emph{The European
  Physical Journal D} \textbf{2007}, \emph{43}, 271--274\relax
\mciteBstWouldAddEndPuncttrue
\mciteSetBstMidEndSepPunct{\mcitedefaultmidpunct}
{\mcitedefaultendpunct}{\mcitedefaultseppunct}\relax
\EndOfBibitem
\bibitem[Latz \latin{et~al.}(2010)Latz, Strautins, and Niedziela]{Corheos2010}
Latz,~A.; Strautins,~U.; Niedziela,~D. \emph{Journal of Non-Newtonian Fluid
  Mechanics} \textbf{2010}, \emph{165}, 764--781\relax
\mciteBstWouldAddEndPuncttrue
\mciteSetBstMidEndSepPunct{\mcitedefaultmidpunct}
{\mcitedefaultendpunct}{\mcitedefaultseppunct}\relax
\EndOfBibitem
\bibitem[Schmidt \latin{et~al.}(2013)Schmidt, Niedziela, Steiner, and
  Zausch]{Corheos2013}
Schmidt,~S.; Niedziela,~D.; Steiner,~K.; Zausch,~J. CoRheoS: Multiphysics
  solver framework and simulation infrastructure for complex rheologies.
  Proceedings NAFEMS World Congress, Salzburg. 2013\relax
\mciteBstWouldAddEndPuncttrue
\mciteSetBstMidEndSepPunct{\mcitedefaultmidpunct}
{\mcitedefaultendpunct}{\mcitedefaultseppunct}\relax
\EndOfBibitem
\bibitem[Novotny and Hecht(2009)Novotny, and Hecht]{Novotny2009}
Novotny,~L.; Hecht,~B. \emph{Principles of Nano-Optics}; Cambridge University
  Press, 2009\relax
\mciteBstWouldAddEndPuncttrue
\mciteSetBstMidEndSepPunct{\mcitedefaultmidpunct}
{\mcitedefaultendpunct}{\mcitedefaultseppunct}\relax
\EndOfBibitem
\bibitem[Agate \latin{et~al.}(2004)Agate, Brown, Sibbett, and
  Dholakia]{Agate04}
Agate,~B.; Brown,~C. T.~A.; Sibbett,~W.; Dholakia,~K. \emph{Opt. Express}
  \textbf{2004}, \emph{12}, 3011--3017\relax
\mciteBstWouldAddEndPuncttrue
\mciteSetBstMidEndSepPunct{\mcitedefaultmidpunct}
{\mcitedefaultendpunct}{\mcitedefaultseppunct}\relax
\EndOfBibitem
\bibitem[Shane \latin{et~al.}(2010)Shane, Mazilu, Lee, and Dholakia]{Shane2010}
Shane,~J.~C.; Mazilu,~M.; Lee,~W.~M.; Dholakia,~K. \emph{Optics Express}
  \textbf{2010}, \emph{18}, 7554\relax
\mciteBstWouldAddEndPuncttrue
\mciteSetBstMidEndSepPunct{\mcitedefaultmidpunct}
{\mcitedefaultendpunct}{\mcitedefaultseppunct}\relax
\EndOfBibitem
\end{mcitethebibliography}
\end{document}